\documentclass[lettersize,journal]{IEEEtran}
\usepackage{amsmath,amsfonts}
\usepackage{algorithmic}
\usepackage{algorithm}
\usepackage{array}
\usepackage[caption=false,font=normalsize,labelfont=sf,textfont=sf]{subfig}
\usepackage{textcomp}
\usepackage{stfloats}
\usepackage{url}
\usepackage{verbatim}
\usepackage{multirow}
\usepackage{xcolor}
\usepackage{etoolbox}
\usepackage{graphicx}
\usepackage{cite}
\usepackage{threeparttable}
\usepackage{orcidlink}
\usepackage{lipsum}
\usepackage{fancyhdr}
\fancypagestyle{sec}{\lhead{\tmpx}}

\fancypagestyle{arXiv}
{
   \fancyhf{}
   \chead{\textcolor{gray}{This article will be presented at the IEEE MTT-S International Microwave Symposium (IMS 2025), San Francisco, CA, USA, USA 15–20, 2025 and accepted to the IEEE Microwave and Wireless Technology Letters (MWTL).}}
   \cfoot{ \fontsize{=9pt}{10pt}\selectfont  \textcolor{gray}{© 2025 IEEE. Personal use of this material is permitted. Permission from IEEE must be obtained for all other uses, in any current or future media, including reprinting/republishing this material for advertising or promotional purposes, creating new collective works, for resale or redistribution to servers or lists, or reuse of any copyrighted component of this work in other works}\\\fontsize{=10pt}{12pt}\selectfont\thepage}
   
}

\makeatletter
\patchcmd{\@maketitle}
  {\addvspace{0.5\baselineskip}\egroup}
  {\addvspace{-0.5\baselineskip}\egroup}
  {}
  {}
\makeatother

\begin{document}

\title{DeltaDPD: Exploiting Dynamic Temporal Sparsity in Recurrent Neural Networks for Energy-Efficient Wideband Digital Predistortion}

\author{ Yizhuo~Wu\orcidlink{0009-0009-5087-7349},
    Yi~Zhu,
    Kun~Qian,
    Qinyu~Chen\orcidlink{0009-0005-9480-6164},
    Anding~Zhu\orcidlink{0000-0002-8911-0905},\\
    John~Gajadharsing,
    Leo~C.~N.~de~Vreede\orcidlink{0000-0002-5834-5461},
    Chang~Gao\orcidlink{0000-0002-3284-4078}
\thanks{Corresponding author: Chang Gao (chang.gao@tudelft.nl)}
\thanks{Yizhuo Wu, Kun Qian, Leo C. N. de Vreede, and Chang Gao are with the Department of Microelectronics, Delft University of Technology, The Netherlands.}
\thanks{Yi Zhu and John Gajadharsing are with Ampleon Netherlands B.V., The Netherlands.}
\thanks{Qinyu Chen is with the Leiden Institute of Advanced Computer Science (LIACS), Leiden University, The Netherlands.}
\thanks{Anding Zhu is with the School of Electrical and Electronic Engineering, University College Dublin, Ireland.}
\thanks{This article was presented at the IEEE MTT-S International Microwave Symposium (IMS 2025), San Francisco, CA, USA, USA 15–20, 2025}}



\maketitle

\begin{abstract}
Digital Predistortion (DPD) is a popular technique to enhance signal quality in wideband RF power amplifiers (PAs). With increasing bandwidth and data rates, DPD faces significant energy consumption challenges during deployment, contrasting with its efficiency goals. State-of-the-art DPD models rely on recurrent neural networks (RNN), whose computational complexity hinders system efficiency. This paper introduces DeltaDPD, exploring the dynamic temporal sparsity of input signals and neuronal hidden states in RNNs for energy-efficient DPD, reducing arithmetic operations and memory accesses while preserving satisfactory linearization performance. Applying a TM3.1a 200MHz-BW 256-QAM OFDM signal to a 3.5 GHz GaN Doherty RF PA, DeltaDPD achieves -50.03\,dBc in Adjacent Channel Power Ratio (ACPR), -37.22\,dB in Normalized Mean Square Error (NMSE) and -38.52\,dBc in Error Vector Magnitude (EVM) with 52\% temporal sparsity, leading to a 1.8$\times$ reduction in estimated inference power. The DeltaDPD code will be released after formal publication at \textcolor{red}{\url{https://www.opendpd.com}}.
\end{abstract}


\section{Introduction}
\thispagestyle{arXiv}
\IEEEPARstart{D}{igital} pre-distortion (\textbf{DPD}) is a popular method to linearize wideband Radio Frequency (\textbf{RF}) Power Amplifiers (\textbf{PA}). Nevertheless, in modern radio digital backends, DPD consumes a substantial portion of power~\cite{wesemann2023energy}. This issue could be further intensified by the potential incorporation of Machine Learning (\textbf{ML}) techniques, such as Recurrent Neural Networks (\textbf{RNNs}), which, despite their promising capabilities, increase power requirements.

Recent progress in ML-based long-term RNN-based DPD for wideband PAs is detailed in~\cite{VDLSTM,PGJANET,DVRJANET,PNRNN}. However, the considerable computational complexity and memory needs of RNN-based DPD systems present major challenges to their efficient implementation in digital signal processing processors for wideband transmitters. This is especially relevant for upcoming 5.5G/6G base stations or Wi-Fi 7 routers, where limited power resources restrict real-time DPD model computation.

Previous methods to tackle DPD energy consumption include reducing the sample rate~\cite{Li2020SampleRate}, utilizing a sub-Nyquist feedback receiver in the observation path~\cite{Hammler2019}, dynamically modifying model cross-terms based on input signal properties~\cite{Li2022}, creating simplified computational pathways for DPD algorithms~\cite{Beikmirza2023}, pruning the unimportant weight of fully connected layer to induce static spatial weight sparsity~\cite{Liu2022} and reducing the precision of DPD models~\cite{Wu2024IMS}. 
\begin{figure}[!t]
    \centering
    \includegraphics[width=\linewidth]{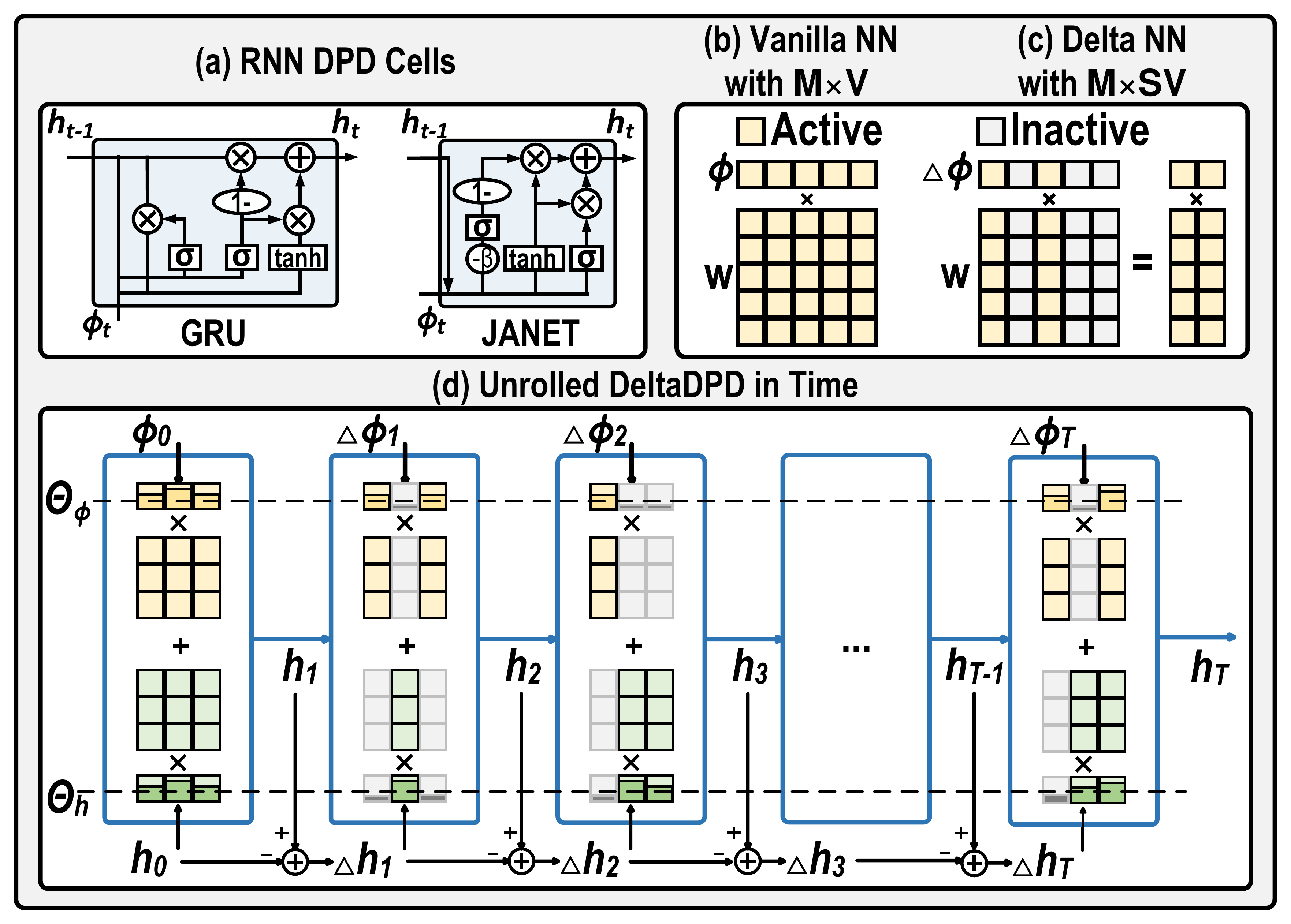}
    \caption{(a) GRU and JANET cell structure with inputs $\phi$ and hidden states $h$; (b) Vanilla network and (c) Delta network M$\times$V; (d) Unrolled DeltaDPD network M$\times$V in time.}
    \label{fig:delta}
\end{figure}

This paper proposes a novel power-saving approach for wideband DPD by inducing and exploiting dynamic temporal sparsity~\cite{liu2024dynamic} in RNN inputs and hidden states using the delta network algorithm~\cite{neil2017delta}. The proposed algorithm decreases both memory access and arithmetic operations by deactivating part of multiplication-and-accumulation (\textbf{MAC}) operations.
It facilitates the design of power-area-efficient RNN computing hardware suitable for DPD deployment in resource-constrained environments. The proposed method can potentially be applied to various RNN-based DPDs.
\section{The DeltaDPD Algorithm}
In this work, we use JANET~\cite{van2018unreasonable} and GRU~\cite{Cho2014GRU} cells, as shown in Figure~\ref{fig:delta}(a), which were adopted in recent DPD studies~\cite{PGJANET,DVRJANET,Wu2024IMS}, to verify the effectiveness of the DeltaDPD in reducing power without significant linearization loss and its adaptability to different RNN architectures. Both the JANET and GRU cells are cascaded with a fully connected layer with 2 output neurons as the output layer.
\vspace{-12pt}
\subsection{The Delta Updating Rule}
Neural networks (\textbf{NNs}) use dense-matrix-dense-vector multiplication (\textbf{M$\times$V}) as illustrated in Figure~\ref{fig:delta}(b). When processing continuous sequential signals using NNs, input data samples $\phi$ and hidden states $h$ of the network could have high autocorrelation, causing small changes ($\Delta$) between neighboring time steps at durations when the derivative of data is low, leading to temporal sparsity in delta input $\Delta\phi$ and delta hidden state vectors $\Delta h$, which can be used to convert M$\times$V into dense-matrix-sparse-vector multiplication (\textbf{M$\times$SV}). 
As depicted in Figure~\ref{fig:delta}(d), by defining thresholds $\Theta_{\phi}$ and $\Theta_{h}$, DeltaDPD skips MAC operations and memory access involving below-threshold $\Delta$ vector elements and their corresponding weight columns, where all gray elements are skipped.

A sequential delta \textbf{M$\times$V} can be derived by:
\begin{align}
\mathbf{y}_t &= \mathbf{W} \mathbf{x}_t,\label{eq1} \\
\mathbf{y}_t &= \mathbf{W} \Delta\mathbf{x}_t + \mathbf{y}_{t-1} = \mathbf{W} (\mathbf{x}_t - \mathbf{x}_{t-1}) + \mathbf{y}_{t-1}, \label{eq2}
\end{align}
where $x$ can be either the RNN input $\boldsymbol{\phi}_t$ or hidden state $\mathbf{h}_{t}$ vector at time $t$, $\mathbf{W}$ represents the weight matrices, and $\mathbf{y}_{t-1}$ is M$\times$V result from the previous time step $t-1$. In Eq.~\ref{eq1}, $\mathbf{W} \Delta\mathbf{x}_t$ becomes \textbf{M$\times$SV} if only computations corresponding to above-threshold $\Delta\mathbf{x}_t$ elements are kept, as given by:
\begin{align}
\Delta\mathbf{x}_t &= \begin{cases}
\mathbf{x}_t - \tilde{\mathbf{x}}_{t-1}, & |\mathbf{x}_t - \tilde{\mathbf{x}}_{t-1}| > \Theta_{x}, \\
0, & |\mathbf{x}_t - \tilde{\mathbf{x}}_{t-1}| \leq \Theta_{x}, \label{eq4}
\end{cases}
\end{align}
\noindent where a piece of memory $\tilde{\mathbf{x}}$ is used to buffer the previous state. To prevent error accumulation over time by memorizing only the last significant change, each $k$-th scalar element $\tilde{x}^{k}$ of vector $\tilde{\mathbf{x}}$ only gets updated when the corresponding $\Delta x^{k}$ exceeds the threshold. This updating rule is defined by:
\begin{align}
\tilde{x}^{k}_{t-1} &= \begin{cases}
x^{k}_{t-1}, & | x^{k}_t - \tilde{x}^{k}_{t-1}| > \Theta_{x}, \\
\tilde{x}^{k}_{t-2}, & |x^{k}_t - \tilde{x}^{k}_{t-1}| \leq \Theta_{x}, \label{eq5}
\end{cases}
\end{align}

\subsection{Definition of DeltaDPD}
Taking the classic GRU-RNN as an example, the pre-activation accumulation of DeltaGRU with input feature $\boldsymbol{\phi}_t = \begin{bmatrix} I_t,\,Q_t,\,|x_t|,\,|x_t|^3,\,sin\theta_t,\,cos\theta_t  \end{bmatrix}$ can be derived by transforming the original GRU equations into their delta forms by following Eqs.~\ref{eq1}$\sim$\ref{eq2}:
\begin{align}
\mathbf{M}_{r,t} &= \mathbf{W}_{ir} \Delta\boldsymbol{\phi}_t + \mathbf{W}_{hr} \Delta\mathbf{h}_{t-1} + \mathbf{M}_{r,t-1}, \label{eq6}\\
\mathbf{M}_{z,t} &= \mathbf{W}_{iz} \Delta\boldsymbol{\phi}_t + \mathbf{W}_{hz} \Delta\mathbf{h}_{t-1} + \mathbf{M}_{z,t-1}, \label{eq7}\\
\mathbf{M}_{n\phi,t} &= \mathbf{W}_{in} \Delta\boldsymbol{\phi}_t + \mathbf{M}_{n\phi,t-1}, \label{eq8}\\
\mathbf{M}_{nh,t} &= \mathbf{W}_{hn} \Delta\mathbf{h}_{t-1} + \mathbf{M}_{nh,t-1}, \label{eq9}
\end{align}
The terms $M_{r}$, $M_{z}$, $M_{n}$ are the pre-activation accumulation of DeltaGRU's reset gate $r$, update gate $z$ and new gate $n$, initialized by the biases of gates $\mathbf{M}_{r,0} = \mathbf{b}_{ir}$, $\mathbf{M}_{z,0} = \mathbf{b}_{iz}$, $\mathbf{M}_{n\phi,0} = \mathbf{b}_{in}$, and $\mathbf{M}_{nh,0} = \mathbf{b}_{hn}$. Therefore, the DeltaGRU-based DPD is defined as:
\begin{align}
\mathbf{r}_t &= \sigma\left(\mathbf{M}_{r,t}\right), \\
\mathbf{z}_t &= \sigma\left(\mathbf{M}_{z,t}\right), \\
\mathbf{n}_t &= \tanh\left(\mathbf{M}_{n\phi,t} + \mathbf{r}_t \odot \mathbf{M}_{nh,t}\right), \\
\mathbf{h}_t &= \left(1 - \mathbf{z}_t\right) \odot \mathbf{h}_{t-1} + \mathbf{z}_t \odot \mathbf{n}_t
\end{align}
 The predicted DPD outputs are generated by a final fully-connected (\textbf{FC}) layer:
\begin{align}
\begin{bmatrix} \hat{I}_t,\,\hat{Q}_t  \end{bmatrix}=\mathbf{\hat{y}}_{t} &= \mathbf{W}_{\hat{y}}\boldsymbol{h}_{t} + \mathbf{b}_{\hat{y}}
\end{align}
The same process can be used to convert the JANET algorithm into a DeltaJANET-based DPD. The delta updating rules of DeltaGRU and DeltaJANET both follow Eqs.~\ref{eq4} and~\ref{eq5}.
\subsection{Theoretical Operation and Memory Access Savings}
In DeltaGRU DPD, the arithmetic operations and memory accesses are dominated by the M$\times$V in Eqs.~\ref{eq6}$\sim$\ref{eq9}. By further considering the overhead in Eqs.~\ref{eq4} and~\ref{eq5} an assuming all vectors have length $n$ and the weight matrices have dimensions $n \times n$, the dense/sparse computational cost $C_{\text{comp}}$ and memorial cost $C_{\text{mem}}$ for calculating M$\times$V and M$\times$SV are given as:
\begin{align}
C_{\text{comp,dense}} &= n^2,\\
C_{\text{comp,sparse}} &= (1 - \Gamma) n^2 + 2n,\\
C_{\text{mem,dense}} &= n^2 + n, \\
C_{\text{mem,sparse}} &= (1 - \Gamma) n^2 + 4n 
\end{align}
\noindent where $\Gamma$ is the overall temporal sparsity by considering zeros in both $\Delta\boldsymbol{\phi}$ and $\Delta\mathbf{h}$. Therefore, the theoretical computation speedup and memory access reduction of a DeltaDPD are approximated as:
\begin{align}
\text{Speedup} &\approx \frac{n}{(1-\Gamma)n+2}, \label{eq20}\\
\text{Memory Access Reduction} &\approx \frac{n+1}{(1-\Gamma)n+4} \label{eq21}
\end{align}

In RNN-based DPD tasks with 500 to 1000 parameters, the value of n for an RNN structure typically ranges from 8 to 20. For example, in DeltaGRU-1067, n equals 15. Considering the overhead terms in Eqs. 15 and 17, only sparsity greater than 27\%  can lead to useful memory access reduction larger than 1 (Eq.19).
Although we give the complete Eq.~\ref{eq20} and~\ref{eq21}, for easy comparison and presentation of results, we estimate the number of active parameters during DeltaDPD inference by:
\begin{equation}
\begin{aligned}
\#\text{Active Params} &= \#\text{DeltaGRU Params} \times \Gamma \\ 
                           &+ \#\text{FC Params}
\end{aligned}
\end{equation}

\begin{figure}[!t]
    \centering
    \includegraphics[width=\linewidth]{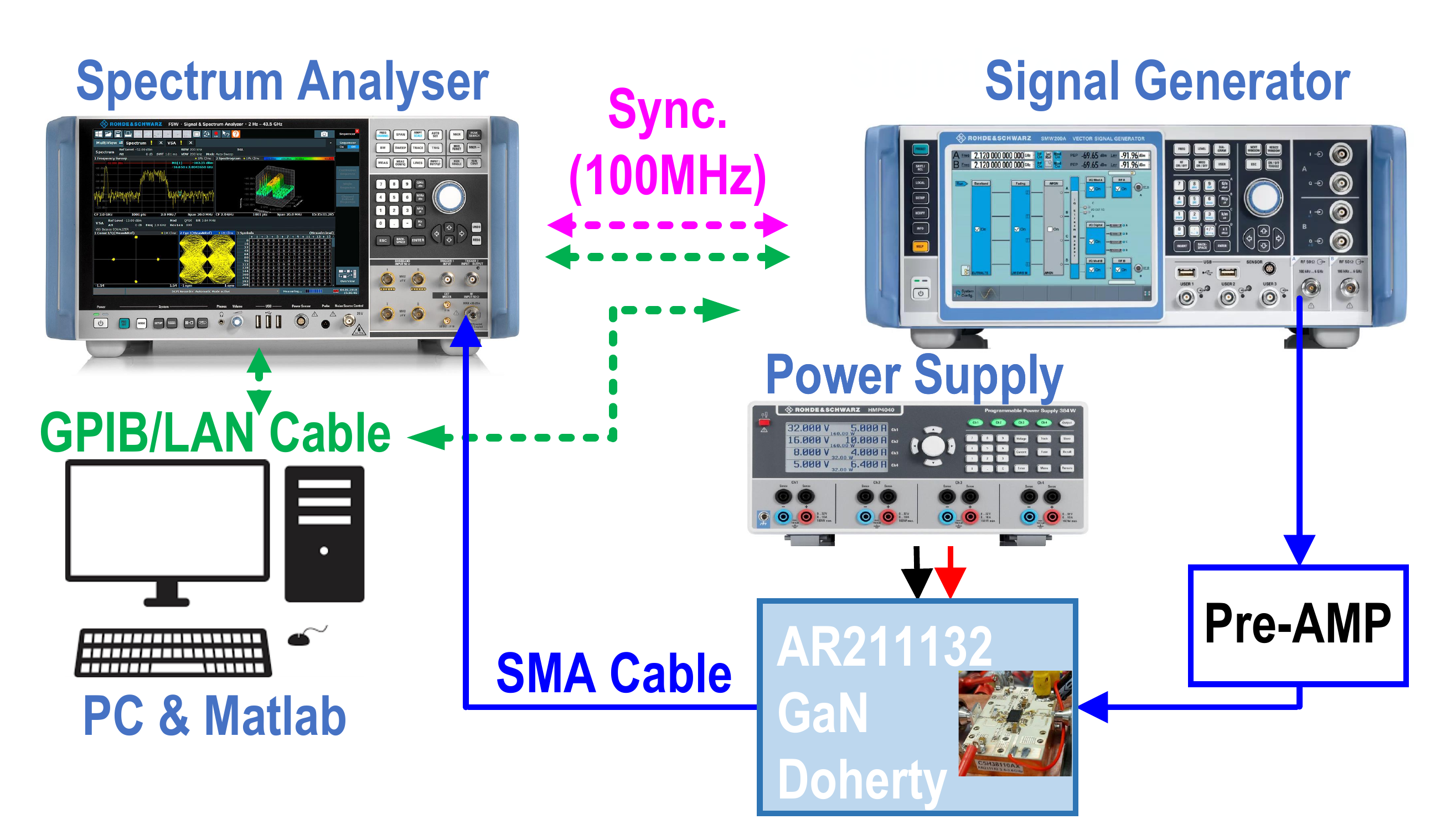}
    \caption{Setup for dataset acquisition and DPD performance measurement.}
    \label{fig:platform}
    \vspace{-13pt}
\end{figure}
\begin{figure}[t]
\centering\includegraphics[width=1.0\linewidth]{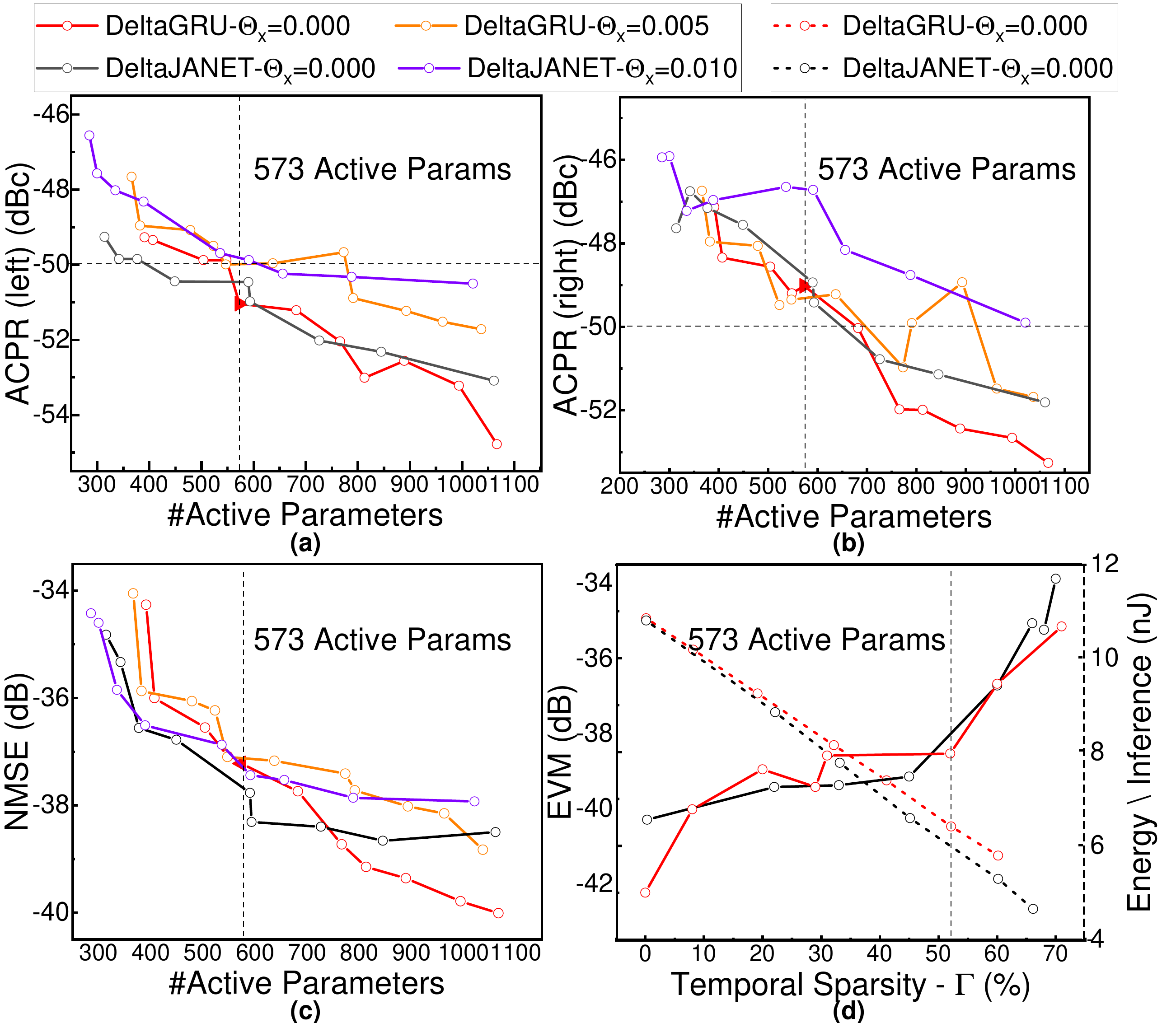}
    \caption{Activated Parameter scan of DPD models vs. (a) ACPR (left) (b) ACPR (right). (c) NMSE (d) Sparsity of 1067-parameter-GRU vs. EVM (left Y-axis) and estimated dynamic power (right Y-axis);}
    \label{fig:results}
    \vspace{-8pt}
\end{figure}
\begin{table*}[t]
\caption{Linearization performance of different DPD models evaluated with TM3.1a 200-MHz 5-channel $\times$ 40-MHz 256-QAM OFDM signals sampled at 983.04 MHz alongside their estimated dynamic power consumption in 7 nm with FP32 parameter precision~\cite{jouppi2021ten}.}
\label{tab:performance}
\resizebox{\linewidth}{!}{%
\begin{threeparttable}
\begin{tabular}{|c|c|c|c|c|c|c|c|c|c|c|}
\hline
\textbf{Class} & \textbf{DPD Models} & \textbf{$\Theta_h$} & \textbf{\begin{tabular}[c]{@{}c@{}}Temporal\\ Sparsity\end{tabular}} & \textbf{\begin{tabular}[c]{@{}c@{}}\#Active\\ Params\end{tabular}} & \textbf{\begin{tabular}[c]{@{}c@{}}NMSE\\ (dB)\end{tabular}} & 
\textbf{\begin{tabular}[c]{@{}c@{}}EVM\tnote{a}\\ (dBc)\end{tabular}} & \textbf{\begin{tabular}[c]{@{}c@{}}ACPR\\ (dBc)\end{tabular}} & 
\textbf{\begin{tabular}[c]{@{}c@{}}Number of \\ MUL/ADD/MEM\end{tabular}} & \textbf{\begin{tabular}[c]{@{}c@{}}Energy/Inference\\ (nJ)\end{tabular}} & \textbf{\begin{tabular}[c]{@{}c@{}}Energy\\ Reduction\end{tabular}} \\ \hline \hline
 & RVTDCNN~\cite{Hu2022RVTDCNN} &  &  & 1007 & -31.64 & -32.43 & -51.75 & 1063/1975/1019 & 9.35 &  \\  
 & PG-JANET~\cite{PGJANET} &  &  & 1130 & -39.77 & -39.94 & -52.91 & 1144/3397/1133 & 10.54 &  \\ 
\multirow{-3}{*}{Prior DPD} & DVR-JANET~\cite{DVRJANET}  & \multirow{-3}{*}{-} & \multirow{-3}{*}{-} & 1097 & -38.02 & -38.24 & -53.79 & 1111/2464/1100 & 10.10 & \multirow{-3}{*}{-} \\ \hline \hline
 & DeltaGRU-1067 & 0 & 0\% & 1067 & -40.01& -42.23 & -54.02 & 1083/2499/1204 & 10.85 & 1x \\ 
 & DeltaGRU-889 & 0.008 & 20\% & 889 & -39.36& -38.95 & -52.50 & 905/2321/1026 & 9.25 & 1.2x \\ 
 & DeltaGRU-766 & 0.016 & 31\% & 766 & -38.73& -38.58 & -52.01 & 782/2198/903 & 8.15 & 1.3x \\  
 & {\color[HTML]{CB0000} \textbf{DeltaGRU-573}} & {\color[HTML]{CB0000} \textbf{0.05}} & {\color[HTML]{CB0000} \textbf{52\%}} & {\color[HTML]{CB0000} \textbf{573}} & {\color[HTML]{CB0000} \textbf{-37.22}}& {\color[HTML]{CB0000} \textbf{-38.52}} & {\color[HTML]{CB0000} \textbf{-50.03}} & {\color[HTML]{CB0000} \textbf{589/2005/710}} & {\color[HTML]{CB0000} \textbf{6.41}} & {\color[HTML]{CB0000} \textbf{1.7x}} \\ 
 & DeltaGRU-504 & 0.1 & 60\% & 504 & -36.67& -37.83 & -49.22 & 520/1936/641 & 5.80 & 1.9x \\ 
 & DeltaGRU-391 & 0.4 & 71\% & 391 & -34.26& -35.14 & -48.20 & 407/1823/528 & 4.78 & 2.1x \\ \cline{2-11}  & DeltaJANET-1062 & 0 & 0\% & 1062 & -38.50& -40.29 & -52.45 & 1078/2494/1198 & 10.80 & 1x \\ 
 & DeltaJANET-845 & 0.004 & 22\% & 845 & -38.66& -39.42 & -51.73 & 861/2277/981 & 8.85 & 1.2x \\ 
 & DeltaJANET-725 & 0.008 & 33\% & 725 & -38.40& -39.37 & -51.40 & 741/2157/861 & 7.78 & 1.4x \\  
 & {\color[HTML]{CB0000}  \textbf{DeltaJANET-593}} & {\color[HTML]{CB0000}  \textbf{0.012}} & {\color[HTML]{CB0000}  \textbf{45\%}} & {\color[HTML]{CB0000}  \textbf{593}} & {\color[HTML]{CB0000}  \textbf{-38.31}}& {\color[HTML]{CB0000}  \textbf{-39.14}} &{\color[HTML]{CB0000}  \textbf{-50.20}} &{\color[HTML]{CB0000}  \textbf{609/2025/729}} & {\color[HTML]{CB0000}  \textbf{6.59}} &{\color[HTML]{CB0000}  \textbf{1.6x}} \\  
 & DeltaJANET-449 & 0.03 & 60\% & 449 & -36.78& -36.72 & -49.05 & 465/1881/585 & 5.30 & 2.0x \\ 
\multirow{-12}{*}{\textbf{This Work\tnote{b}}} & DeltaJANET-377 & 0.05 & 66\% & 377 & -35.33& -35.06 & -48.54 & 393/1809/513 & 4.65 & 2.3x \\ \hline
\end{tabular}%
\begin{tablenotes}
\item[a] Due to limitations in the experimental setup, the EVM is calculated based on the input signal and the measured output signal rather than the reference grid and the measured output signal. Additionally, the mild CFR applied to the input signal may cause a degradation in the EVM.
\item[b] We use $\Theta_{\phi}=0$ for all DeltaDPDs in this table. 
\end{tablenotes}
\end{threeparttable}
}
\vspace{-12pt}
\end{table*}
\section{Experimental Results}
\subsection{Experimental Setup}
Figure~\ref{fig:platform} illustrates the experimental setup. 
The \texttt{TM3.1a} 5$\times$40-MHz (200-MHz) 256-QAM OFDM baseband I / Q signal with 10.01\,dB Peak-to-Average Power Ratio (\textbf{PAPR}) was emitted by \texttt{R\&S-SMW200A} and amplified by a 3.5GHz GaN Doherty PA at 41.5\,dBm average output power with and without DPD. The output signal was digitized using an \texttt{R\&S-FSW43} analyzer. Since this spectrum analyzer lacks EVM calculation capability, the EVM was determined by comparing the input signal with the digitized output signal instead of using the reference grid. The dataset, comprising 98304 samples, was divided into 60\%,20\% and 20\% for training, validation, and testing.

The end-to-end DPD learning process involves backpropagation through a pre-trained 2751-parameter -40.04\,dB-NMSE DGRU PA behavioral model~\cite{wu2024opendpd} with the newly measured PA dataset. The models were trained for 200 epochs using the ADAMW optimizer with an initial learning rate of 5E-3 with \texttt{ReduceOnPlateau} decay and a batch size of 64.
\subsection{Results and Discussion}

\begin{figure}[t]
    \centering
    \includegraphics[width=0.9\linewidth]{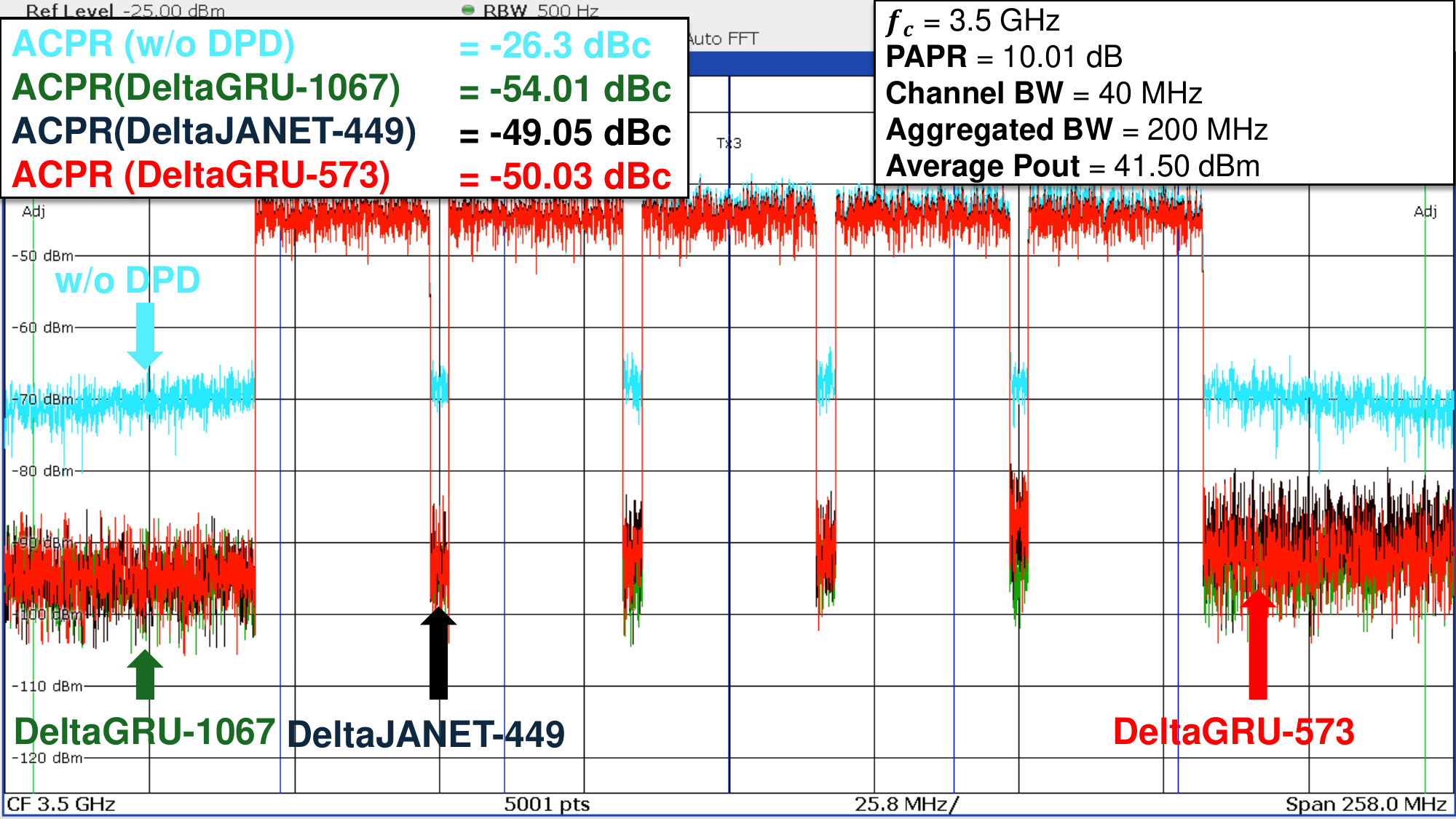}
    \caption{Measured spectrum on the 200\,MHz TM3.1a signal.}
    \label{fig:PSD}
    \vspace{-12pt}
\end{figure}

Table~\ref{tab:performance} compares the NMSE, ACPR, and EVM results for different DPD models alongside the number of MUL, ADD operations, and 8KB SRAM accesses. The estimation method follows~\cite{Wu2024IMS}. The DeltaGRU-573 DPD model with $\Theta_{\phi}$ of 0, $\Theta_h$ of 0.05 achieves an ACPR of -50.03\,dBc, an NMSE of -37.22\,dB and an EVM of -38.52\,dBc while estimated to consume 6.41\,nJ per inference in 7\,nm technology. The DeltaGRU-573 demonstrates the most considerable power reduction while maintaining the ACPR better than -50\,dBc, as highlighted by the horizontal dashed lines in Fig.~\ref{fig:results}. 

Figs.~\ref{fig:results} (a), (b), and (c) show the correlation between ACPR/NMSE and estimated energy/inference against \#active parameters of DeltaGRU/DeltaJANET covering 300 to 1100 active parameters. 
Even at high temporal sparsity of around 70\% with around 400 active parameters, DeltaGRU and DeltaJANET still maintain ACPR values better than -48\,dB. Comparing the performance of various $\Theta_{\phi}$, utilizing temporal sparsity of input feature even close to 0 in the DPD task degrades the linearization performance by 1.57\,dB because the DPD performance is highly sensitive to the I/Q sampling rate. Fig.~\ref{fig:results} (d) presents the estimated energy per inference in 7\,nm of DeltaDPDs. The DeltaGRU-573 model realizes a 1.7$\times$ power reduction over the DeltaGRU-1067 network. 

\begin{figure}
    \centering\includegraphics[width=0.95\linewidth]{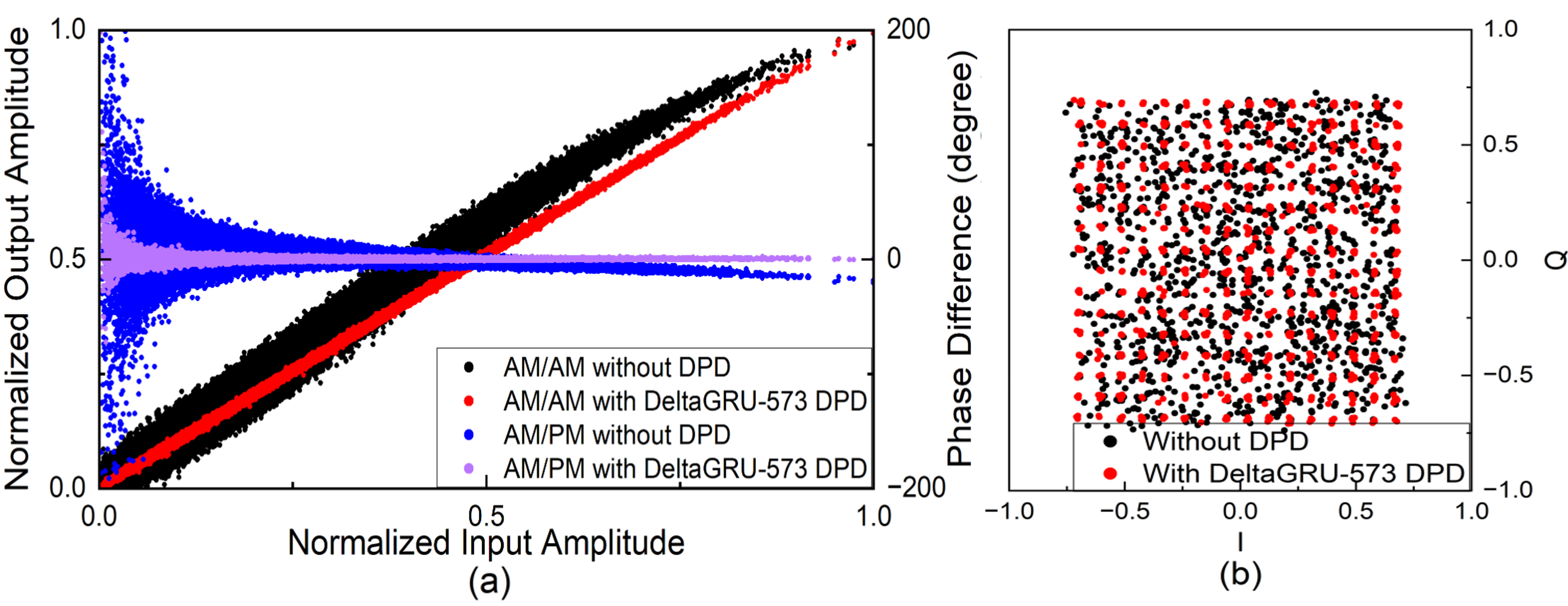}
    \caption{(a) AM/AM and AM/PM characteristics (b) constellation map with and without DPD for the 200-MHz OFDM signal.}
    \label{fig:amam}
    \vspace{-12pt}
\end{figure}

Fig.~\ref{fig:PSD} displays the measured spectrum with and without DPDs, which confirms that the DeltaGRU-573 model achieves ACPR of -50\,dBc. Fig.~\ref{fig:amam} exhibits the AM/AM, AM/PM characteristics and constallation map with and without DPDs. 


\subsection{Comparison to Prior Works}
Due to power constraints in DPD applications, state-of-the-art models are typically limited to around 1000 parameters~\cite{PNRNN}, making NN performance particularly susceptible to compression and sparsity of input compared to delta networks with parameters more than 160000 in other domains ~\cite{liu2024dynamic,neil2017delta}. The previous approaches of lightening the DPD model have primarily relied on static spatial weight pruning static spatial NN weights~\cite{Liu2022,li2024gpu}. Using a 100\,MHz OFDM signal, Liu et al.~\cite{Liu2022} achieved an ACPR of -45.5\,dBc with a pruned convolutional NN-based DPD model containing 106 parameters, reduced from 158. Li et al.~\cite{li2024gpu} demonstrated an ACPR of -45.1\,dBc at 200\,MHz using a pruned phase-normalized time-delay NN with 909 parameters. However, these unstructured pruning methods create irregular distributions of nonzero values in weight matrices, causing unbalanced workloads among hardware arithmetic units and limiting real speedup or efficiency gains in actual hardware implementations. In contrast, our proposed DeltaDPD achieves a superior ACPR of -50.03\,dBc at 200\,MHz with only 573 parameters while maintaining structure.

\section{Conclusion}
This work introduces DeltaDPD, a novel method for energy-efficient RF power amplifier linearization that leverages dynamic temporal sparsity. By reducing computational complexity and memory access compared to conventional approaches, DeltaDPD achieves power savings while maintaining robust linearization performance. 





\bibliographystyle{IEEEtran}

\bibliography{IEEEabrv,ref}

\end{document}